\documentclass[letterpaper,12pt]{revtex4}   
\usepackage[draft]{hyperref} 

\usepackage{epsfig}
\usepackage{graphicx}
\usepackage{subfigure}
\usepackage{amsmath}

\begin{document}


\title{Astrointerferometry with discrete optics}

\author{Stefano Minardi$^{1,2*}$, Thomas Pertsch$^1$}

\address{
$^1$Institute of Applied Physics, Friedrich-Schiller-Universit\"at Jena, Max-Wien-Platz 1,07743 Jena, Germany\\
$^2$Astronomisches Institut und Universitäts Sternwarte, Friedrich-Schiller-Universit\"at Jena, Max-Wien-Platz 1,07743 Jena, Germany\\
$^*$Corresponding author: stefano.minardi@uni-jena.de
}

\begin{abstract}
We propose an innovative scheme exploiting discrete diffraction in a two-dimensional array of coupled waveguides to determine the phase and amplitude of the mutual correlation function between any pair of three telescopes of an astrointerferometer.
\end{abstract}


\maketitle
 
The Van de Cittert theorem allows the determination of the angular shape of an incoherent source from the measurement of the complex spatial correlation function of the emitted light\cite{BornWolf}. 
This feature is routinely used in astrointerferometry to achieve aperture synthesis by the coherent combination of light collected from an array of telescopes aimed at the same astronomical target.
In simple terms, high resolution images of stars are obtained from the Fourier transform of the measured complex correlation function (complex visibilities) which is sampled at frequencies determined by the projected baselines connecting pairs of telescopes.
The current challenge of astronomical interferometry is to enable simultaneous evaluation of the complex visibilities on a large number of baselines \cite{CHARA}, to allow interferometric imaging techniques of fast astronomical events such as exoplanet transits \cite{CHARA_exoplanets} or nova/supernova explosions \cite{Chesneau}. 

To date, most of the existing or planned interferometric facilities are using bulk optics to perform multiple telescope beam combinations. While this solution helps to keep costs relatively low, the setups are typically complex, bulky and require special construction to keep mechanical vibrations to an acceptable level. Due to their inherent thermal and mechanical stability, integrated optic beam combiners \cite{VINCI} could reveal decisive to perform multiple beam combination effectively. Recently, a planar
integrated chip allowing the measurement of the phase over 6 baselines among 4 telescopes has been realized and
tested \cite{LAOG}. 
Due to geometrical constraints, the complexity of planar integrated optical components grows rapidly as more telescopes and baselines are added. Three dimensional geoemtries, such as those enabled by laser-writing techniques \cite{Nolte03,3Dcomponents}, could relax geometrical constraints e.g. by allowing fiber cross-overs, resulting in a design simplification of multiple beam combiners. 

In this paper, we put forward an elegant and innovative scheme allowing the measurement of the spatial coherence function sampled by three telescopes by means of a three dimensional integrated component. The device exploits discrete diffraction effects in a two-dimensional array of coupled waveguides to perform a two-quadrature measurement of the phase delay between all possible baselines in a 3-telescope stellar interferometer (see scheme in Fig. 1(a)).

Starting point of our work is the consideration that free space diffraction entails intrinsically the capability of
measuring the wavefront profile. Phase retrieval from Gabor-type holograms (also known as shadowgraphy) is
probably the most striking example of exploitation of diffraction for phase metrology purposes \cite{Allen01,Gopal07}. Light propagation in an array of evanescently coupled waveguides is similar to conventional diffraction, however it is
bound to a system with finite degrees of freedom. 

Here we demonstrate that the measurement of the discrete
diffraction intensity pattern at the end of a N=3x3 waveguide array can be uniquely related to the phase and amplitude
of the mutual correlation function of three monochromatic fields coupled to suitable input waveguides. 
We chose to number of the sites of the 2D array
of waveguides progressively from left-to-right and from top-to-bottom. Let $A_\textrm{i}$, $A_\textrm{j}$ and $A_\textrm{k}$ be
the complex amplitude of the light of the three beams at the input face of the waveguide array and injected in the
sites $\textrm{i}$, $\textrm{j}$ and $\textrm{k}$, respectively. The field at the output of the $\mathrm{n}^{th}$ waveguide of the array is a linear combination of the input fields:
\begin{equation}
U_\textrm{n} = a_\textrm{n,i}A_\textrm{i} + a_\textrm{n,j}A_\textrm{j} + a_\textrm{n,k}A_\textrm{k}. 
\end{equation}
The coefficients $a_\textrm{n,i}$, $a_\textrm{n,j}$ and $a_\textrm{n,k}$ are a function of $z$ and can be calculated exactly \cite{Szameit07} or obtained by integration of the field propagation equations over the sample length $L$:
\begin{equation}
\frac{\partial U_\mathrm{k}}{\partial z}=-i\frac{\pi}{2 L_c}\sum_\mathrm{l}^N C_{\mathrm{k,l}}U_\mathrm{l},
\end{equation}
where $L_c$ is the coupling length of the fiber arrays \cite{Szameit07} and $C_{\mathrm{k,l}}$ is the matrix describing the relative strength of the coupling between neighbouring waveguides.
The resulting intensity pattern $I_\textrm{n} =< |U_\textrm{n}|^2>$ can be cast in the following matrix form:
\begin{equation}
I_\mathrm{n}=\sum_{\mathrm{k}=1}^N \alpha_{\mathrm{n},\mathrm{k}} J_\mathrm{k},
\end{equation}
provided that we define the $J_k$ as:
\begin{eqnarray}
J_\mathrm{1}=<|A_\mathrm{i}|^2>, \quad J_\mathrm{2}=<|A_\mathrm{j}|^2>, \quad J_\mathrm{3}=<|A_\mathrm{k}|^2>\nonumber\\
J_\mathrm{4}+iJ_\mathrm{5}=<A_\mathrm{i}A_\mathrm{j}^*>=\Gamma_{\mathrm{ij}} \quad J_\mathrm{4}-iJ_\mathrm{5}=<A_\mathrm{j}A_\mathrm{i}^*>=\Gamma_{\mathrm{ij}}^*\\
J_\mathrm{6}+iJ_\mathrm{7}=<A_\mathrm{i}A_\mathrm{k}^*>=\Gamma_{\mathrm{ik}} \quad J_\mathrm{6}-iJ_\mathrm{7}=<A_\mathrm{k}A_\mathrm{i}^*>=\Gamma_{\mathrm{ik}}^*\nonumber\\
J_\mathrm{8}+iJ_\mathrm{9}=<A_\mathrm{j}A_\mathrm{k}^*>=\Gamma_{\mathrm{jk}} \quad J_\mathrm{8}-iJ_\mathrm{9}=<A_\mathrm{k}A_\mathrm{j}^*>=\Gamma_{\mathrm{jk}}^*\nonumber
\end{eqnarray}
Here the functions $\Gamma = |\Gamma|\exp(i\Phi)$ are the complex field correlation functions of all possible pairs of incoming fields \cite{BornWolf}. In particular, it is easy to recognize that the variables $J_4$, $J_6$ and $J_8$ represent the real part of the corresponding correlation functions, while $J_5$, $J_7$ and $J_9$ are their imaginary parts.
The exact retrieval of the mutual coherence of the input fields is possible whenever the matrix $\alpha_{n,k}$ is invertible and well conditioned. We numerically calculated the matrices for all possible combinations of three input fields on a 3x3 square array of waveguides of varying length. The optimal input combination and array length allow obtaining a maximum error of 10\% on the retrieved correlation function components $\{J_n\}$ with a 1\% precision measuerement of the output intensity fields $\{I_n\}$. Figure 2 shows a possible output of such a configuration for three fully coherent injected fields exploiting the indicated phase differences.

Better insight in the potential of our interferometric scheme is given by the results of the simulated observation of a binary star. Figure 3(a) shows the amplitude of the spatial correlation function expected for the binary star depicted in the inset, and observed by the array of Figure 1. The traces of the sampled points during 6 hours of observation with the astrointerferometer are superimposed to the image. For simplicity, it has been assumed that the star is located along the direction of the rotation axis of the Earth, and the telescope array is located at the pole where the star is observable. 
For the chosen baselines, the amplitude of the visibilities and the closure phase (sum of the phases of the three correlation functions \cite{Rogstad68}) during the observation time are shown in Figure 3(b-c), along with the estimated values and measurement errors obtained from the realistic modeling of the discrete beam combiner. 
In the model, we calculate for the selected observation points the expected intensities at the output of the discrete beam combiner $\{I_n\}$. We then model the error in the intensity measurement by adding to each $\{I_n\}$ a pseudo-random number representing the photon shot noise of detected intensity. We assume that the detector has a minimum shot noise of 0.5\%, which is typical for 16 bit detectors working near the saturation level, which we set equal to the maximum value of the $\{I_n\}$. For each of the 100 noise realizations we calculate the retrieved values of $\{J_n\}$. These sets are then used to calculate the average value and the standard deviation of the amplitude of the visibilities and the closure phases. 
  
The simulations point out that the proposed beam combiner could reach a precision of the normalized visibility measurement of $\pm 0.05$, while the the closure phase could be determined with a error below $\pm 0.5$ degree.
These values are in line with the current performance of existing interferometers, we point out that the performance of the proposed wave coherence meter can be improved, \textit{e.g.} by the optimization of the geometry of the fiber array.
  
In conclusion, we have demonstrated theoretically that a two dimensional array of coupled waveguides can be used to determine simultaneously the amplitude and the phase of the correlation function of a light source, sampled with an array of 3 apertures (telescopes). The device could be realized by means of laser writing techniques and can be virtually scaled up to an arbitrary number of telescopes and baselines. Applications to both fringe tracking and visibility measurement of astronomical targets are foreseen. Moreover, because of the small dimensions and lightness of integrate optics, beam combination by means of discrete optics could result attractive in future space-borne astronomical interferometers \cite{darwin}.

\clearpage

\section*{List of Figure Captions}

Figure 1. A pictorial view of the concept of the integrated discrete beam combiner. The light collected by three telescopes arrenged as in (a) is injected in three selected sites of a two dimensional array of coupled waveguides ((b) Left). The resulting discrete diffraction intensity pattern at the end of the array ((b) Right) gives unique information about the amplitude and phases of the mutual correlation function of the observed star along the three baselines of the telescope array.

\noindent Figure 2. Example of an output resulting from the injection of three equally intense, fully coherent beams in the highlighted waveguides of a sample length of $L = 1.8 L_c$ ($L_c$ is the coupling length of the fiber array, for the definition see \cite{Szameit07}). The phase differences between the various input fields are reported at the bottom of the picture.

\noindent Figure 3. Simulation of the observation of a binary star. (a) Amplitude of the spatial correlation function of the light emitted by the binary star depicted in the inset. The traces of the projected baselines of the interferometric array during an observation of 6 hours are superposed to the plot. Parameters of the binay star: speraration = 7 mas, ratio of  intensities = 2.512 (corresponding to 1 magnitude). Retrieved amplitude (b) and closure phase (c) of the correlation function sampled along the trajectories of the projected baselines, as simulated by our model of a noisy discrete coherence meter. The expected values of the correlation function are represented by the continous lines.

\clearpage

\section*{Figures and Tables}

\begin{figure}[htbp]
\centerline{\includegraphics[width=8.3cm]{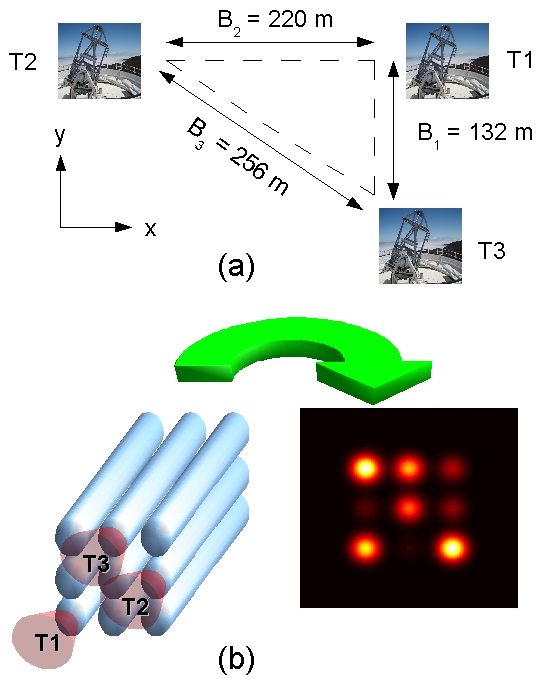}}
\caption{A pictorial view of the concept of the integrated discrete beam combiner. The light collected by three telescopes arrenged as in (a) is injected in three selected sites of a two dimensional array of coupled waveguides ((b) Left). The resulting discrete diffraction intensity pattern at the end of the array ((b) Right) gives unique information about the amplitude and phases of the mutual correlation function of the observed star along the three baselines of the telescope array.}
\end{figure}
\clearpage

\begin{figure}[htbp]
\centerline{\includegraphics[width=8.3cm]{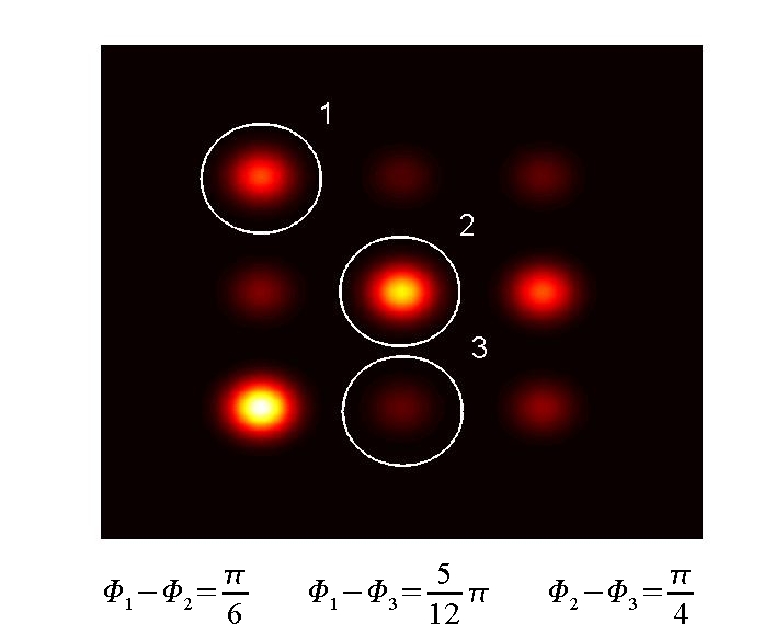}}
\caption{Example of an output resulting from the injection of three equally intense, fully coherent beams in the highlighted waveguides of a sample length of $L = 1.8 L_c$ ($L_c$ is the coupling length of the fiber array, for the definition see \cite{Szameit07}). The phase differences between the various input fields are reported at the bottom of the picture.}
\end{figure}
\clearpage

\begin{figure}[htbp]
\centerline{\includegraphics[width=8.3cm]{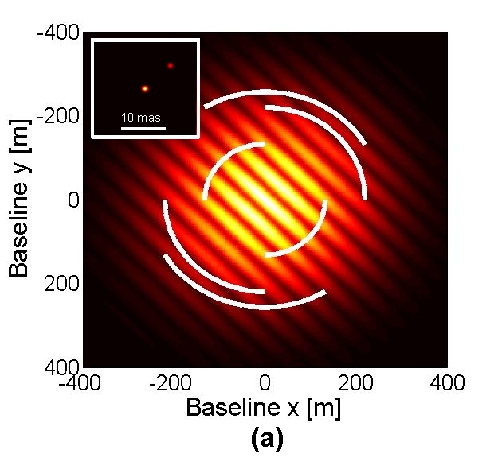}}
\centerline{\includegraphics[width=8.3cm]{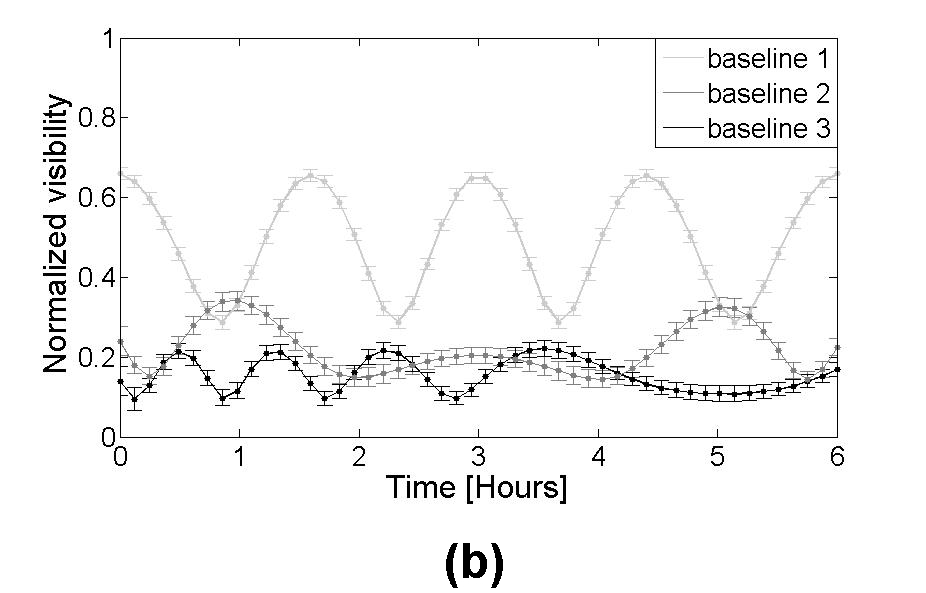}}
\centerline{\includegraphics[width=8.3cm]{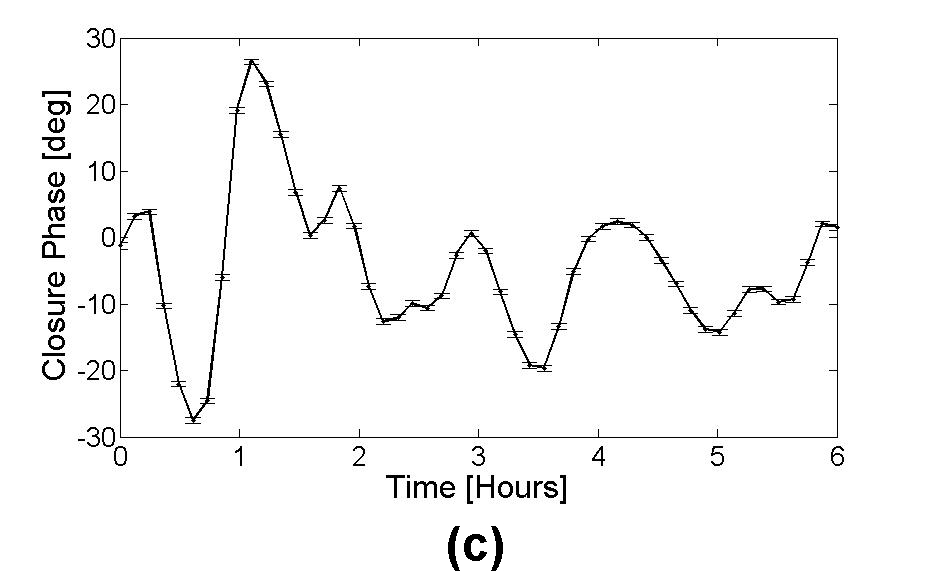}}
\caption{Simulation of the observation of a binary star. (a) Amplitude of the spatial correlation function of the light emitted by the binary star depicted in the inset. The traces of the projected baselines of the interferometric array during an observation of 6 hours are superposed to the plot. Parameters of the binay star: speraration = 7 mas, ratio of  intensities = 2.512 (corresponding to 1 magnitude). Retrieved amplitude (b) and closure phase (c) of the correlation function sampled along the trajectories of the projected baselines, as simulated by our model of a noisy discrete coherence meter. The expected values of the correlation function are represented by the continous lines.}
\end{figure}

\end{document}